# Comparing the Matter and Interactions Curriculum with a Traditional Physics Curriculum: A Think Aloud Study

Keith R Bujak, bujak@gatech.edu[1]
Richard Catrambone, rc7@prism.gatech.edu[1]
Marcos D. Caballero, danny.caballero@physics.gatech.edu[2]
M Jack Marr, mm27@prism.gatech.edu[1]
Michael F Schatz, schatz@cns.physics.gatech.edu[2]
Matthew A Kohlmyer, makohlmy@unity.ncsu.edu[3]

**Abstract**

Physics curricula across the US fail to prepare students adequately to solve problems, especially novel problems. A new curriculum, Matter and Interactions (M&I), was designed to improve student learning by organizing concepts around fundamental principles. Despite this, students taught in traditional classes continue to outperform M&I students on force concept questions. This study aimed to determine the underlying issues related to the performance differential. Students from both courses solved questions from the Force Concept Inventory (FCI) while verbally describing their reasoning. Analysis of the transcripts revealed that M&I students failed to employ the fundamental principles, and traditional students used simple physics facts to help identify the correct answers. Neither of these methods would be sufficient for solving more complex problems.

## Theoretical Framework

Each year in the US, more than 100,000 students take calculus-based introductory physics. These students must obtain a good working knowledge of introductory physics, because physics concepts underpin the content of many advanced science and engineering courses required for the students' degree programs. Unfortunately, rates of failure and withdrawal in these courses are often high, and a large body of research has shown that student misconceptions about physics persist even after instruction has been completed (Chi, 2005; Hake, 1998, Halloun & Hestenes, 1985). Research has shown that many students organize their physics knowledge poorly (Chi, Feltovich & Glaser, 1981; Dufresne, Gerace, Hardiman & Mestre, 1992). In courses with a traditional mechanics curriculum, students commonly structure their understanding around laundry lists of formulas with little understanding of the conceptual underpinnings (Reif, 2008). As a consequence, their problem solving ability is limited.

The Matter and Interactions (M&I) curriculum, developed by Chabay and Sherwood (2007), aims to improve student understanding by organizing mechanics knowledge around three fundamental principles (conservation laws). This design reflects the practice of using a small amount of central knowledge (Figure 1) as the starting point for further expansion by systematically specifying the associations with more detailed subordinate knowledge (Reif, 2008). The more explicit global hierarchical structure of mechanics presented by M&I should be more effective than the standard "many formulas" approach in the traditional curriculum. This improved organization should help M&I students retrieve and use mechanics knowledge more readily and flexibly than their peers in a traditional course.

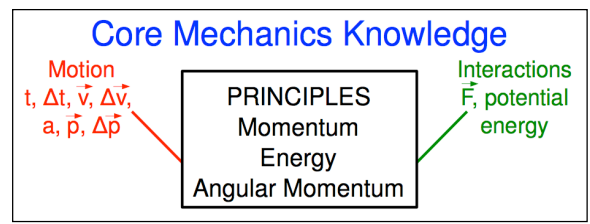

*Figure 1:* Central knowledge about point particle mechanics as taught by the M&I curriculum. (Reif, 2008)

The Force Concept Inventory (FCI) is a standardized multiple-choice test that has gained widespread acceptance in the physics community (Hake, 1998; Hestenes, Wells, & Swackhamer, 1992). At Georgia Tech in the fall of 2006, this instrument showed that students in the traditional course, on average, achieved a normalized gain (i.e., the ratio of the difference between pre- and post-test scores to the total possible increase in score) of 0.36 ± 0.02, while their counterparts in the M&I course achieved normalized gains of only 0.23 ± 0.03.

---

[1] School of Psychology, Georgia Institute of Technology
[2] School of Physics, Georgia Institute of Technology
[3] Department of Physics, North Carolina State University

## Objective

The present study was designed to examine the sources of the performance differences between students of the traditional and M&I courses.

## Method and Procedure

Thirty-four Georgia Tech undergraduates participated in the study. Participants previously completed either the traditional (n = 20) or M&I (n = 14) introductory physics course, receiving a grade of A, B or C (one student's grade was not available). A $X^2$ test on grade distributions yielded no differences between the two conditions, $X^2(3, N = 34) = 1.78$, $p = .62$. All but two of the participants last took the course in the fall, summer or spring semesters immediately preceding the experiment; the other two took the course two years prior (one in traditional, one in M&I). For textbooks, the traditional course worked with *Physics for Scientists and Engineers: A Strategic Approach* (Knight, 2007) and the M&I course used *Matter and Interactions Vol. 1: Modern Mechanics* (Chabay & Sherwood, 2007).

Participants solved 10 physics problems selected from the FCI while describing their reasoning. Though the complete test consists of 30 questions, 10 were chosen for this study based on pilot testing to ensure they covered the proper range of topics and to reduce fatigue during the think aloud protocol. Prior to working on the problems, participants completed a warm-up activity (i.e., a game of tic-tac-toe) to gain experience with the think aloud procedure. Participants were explicitly instructed to explain *why* they decided upon each move.

Upon completing the warm-up, the participant worked at a self-guided pace to solve the 10 FCI problems while describing what they were doing and why. In the case that the participant failed to speak for approximately 10 seconds, the researcher prompted the participant to talk about what he or she was thinking. Audio and video recording captured the participants' problem solving processes for subsequent transcription and analysis.

## Results

The null hypothesis was that the traditional and M&I groups would respond with similar accuracy. An independent samples *t*-test was conducted to compare overall performance. There was no significant difference in the overall scores on the 10 problems solved by traditional (M = 6.85, SD = 2.50) and M&I (M = 5.57, SD = 2.03) students, $t(32) = -1.58$, $p = .12$. To assess the differences for each of the 10 questions, a set of $X^2$ tests were employed. The only question that showed a significant difference between the groups was question 8 (the falling balls problem): traditional students outperformed M&I students, $X^2(1, N = 34) = .026$, $p = .042$. See *Figure 3* for an overview of the results.

The researchers selected two questions for detailed analysis of the transcripts. Question 2 from this think aloud—which is question 17 on the FCI[4]—was selected given that the difference between the two curricula was quite large, albeit non-significant. In this problem, students are asked to identify the individual forces acting on an elevator that is moving up a shaft at a constant speed. Question 8 from this think aloud—which is question 1 on the FCI—was selected because of the significant difference in performance between groups and because it involved a different set of underlying physics principles. In this problem, students are asked to determine the relative time it takes for two balls of different masses to reach the ground when dropped. Quantitative and qualitative data are presented for both questions. Question 5 and 9, which also had large (but non-significant) group differences, were not selected as they were part of multiple-question series.

### Question 2: The Moving Elevator Problem

The majority of responses were divided between options A ("the upward force by the cable is greater than the downward force of gravity") and B [the correct answer] ("the upward force by the cable is equal to the downward force of gravity"). The distribution between the two response choices was reversed for the two groups. More traditional students selected the correct response compared to M&I students (60% and 43%, respectively), while more M&I students selected the most popular incorrect answer compared to traditional students (43% and 25%, respectively). The other three options comprised only 15% and 14% of the remaining responses for traditional and M&I, respectively.

The researchers reviewed each transcript for the following statements: (1) correct answer; (2) constant velocity means no acceleration; (3) no acceleration means no net force; (4) constant velocity means no net force; (5) constant velocity means no change in momentum; (6) no change in momentum means no net force; (7) mentions momentum or change in momentum; (8) presence of a net force in the direction of motion [an incorrect comment]; and (9) drew a diagram. Five people coded the problems independently. Of the 306 total codes, 6% resulted in one person initially disagreeing with the other four, and 3% resulted in two disagreeing with the other three. These few disagreements were resolved through discussion.

The M&I curriculum places great emphases on the three core principles of mechanics, the momentum principle being the first presented in the course. Given this, the most surprising finding from our analysis is the near complete absence of the word "momentum" or the phrase "change in momentum" by M&I students. Only one of the 14 M&I participants mentioned momentum, albeit incorrectly. This participant incorrectly concluded that a constant velocity means zero momentum, when in fact a constant velocity means no *change in* momentum.

---
[4] For more information about the Force Concept Inventory, visit http://modeling.asu.edu/R&E/research.html.

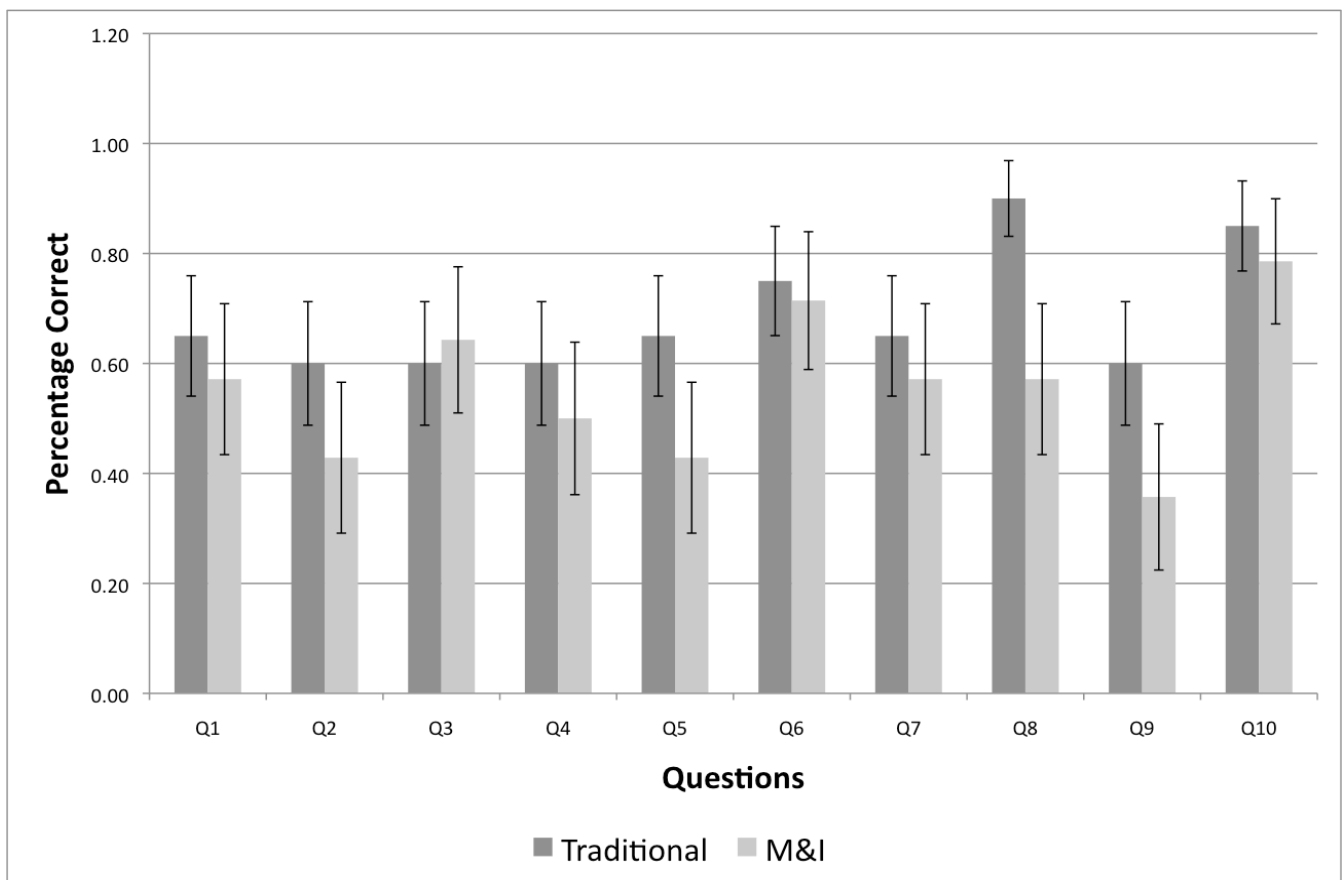

*Figure 3:* Performance of student on FCI question. Error bars represent standard error of the mean.

The incorrect conclusion that there *is* a net force in the upward direction of the motion of the elevator was asserted by 71% of M&I and by 50% of traditional students. The notion of a net force was generally apparent through some implicit statements made by the participants (e.g., "if they [the upward and downward forces] were equal, the elevator wouldn't be moving, and if the force of gravity was greater, then it would be going downwards"). The confusion points at a general failure to understand the proper connection between force and motion (McDermott, et al., 2002).

## Question 8: The Falling Balls Problem

Traditional students significantly out-performed (90% correct) M&I students (57% correct) on this problem. M&I students were largely distracted by options A and D (21% for each). Both of these options show that many of these students believed that heavier objects fall faster, a common misconception among introductory physics students (Hestenes & Swackhamer, 1992). It is true that a larger force would be exerted on the more massive ball, but both balls experience the same acceleration.

The researchers reviewed each transcript for the following statements: (1) correct answer; (2) claim that mass does not matter; (3) used "commonsense" reasoning; (4) stated that acceleration is the same for both; (5) used $F = ma$; (6) concluded that a difference in force on the balls results in a different time to fall [an incorrect comment]; (7) used kinematic equations; (8) mentioned momentum; (9) mentioned air resistance; and (10) drew a diagram. Four people coded the problems independently. Of the 306 total codes, 10% resulted in one person initially disagreeing with the other three, and 6% resulted in two disagreeing with the other two. These few disagreements were resolved through discussion.

One of the most commonly known formulas in physics led M&I students astray: $F = ma$. Five of the 14 M&I students incorrectly stated that the difference in the forces acting on the balls—a correct conclusion—would result in a different amount of time for the balls to reach the ground—an incorrect conclusion. Moreover, all but one of the remaining M&I students (who did *not* take the aforementioned force-based approach) correctly answered the problem. Some of these students simply stated that mass does not matter in this type of problem, or implied that it did not—reflecting a "common" fact that all bodies fall at the same rate under influence of gravity.

Of the 18 traditional students who provided a correct answer, 14 of them simply stated that mass didn't matter or that the answer was common knowledge. It must be noted that only two traditional students used $F = ma$. One student discontinued using $F = ma$ shortly after starting and switched to kinematic equations, and the other superfluously used it to conclude that the acceleration due to gravity is a constant.

## Conclusions

From the detailed analysis of these questions, we conclude that the challenge facing students of the M&I curriculum is properly implementing the principles explored during the course to conceptual force and motion problems. The transcripts from the moving elevator problem expose students' erroneous reasoning in how forces relate to motion, given velocity and acceleration. Moreover, M&I students failed to grasp the notion of net force and the fact that an object moving in a straight-line at constant speed experiences no net force.

The falling balls problem illuminated additional difficulties with the concept of force. Here, students tended to grasp the idea that gravity exerts force on bodies in different ways depending on their masses, but they did not seem to grasp that the motion (i.e., acceleration) is independent of mass during free fall (ignoring affects of air resistance).

Simply having a basic understanding of common ideas in physical systems is sufficient to solve these problems. Students who complete the traditional physics course seem to gain this ability; M&I students appear to fail at this. Even though the M&I course employs a novel curriculum built around sound principles, students do not seem to necessarily gain the visceral understanding of how physical systems behave in a basic way.

The factors controlling incorrect solutions remain somewhat uncertain. Two possibilities are (1) that the students are failing to understand the difference between individual, non-zero forces and *net* force; or (2) they are failing to understand the relationship between velocity and acceleration as they relate to net force.

## Significance

Although the M&I organization should be superior to the traditional organization in introductory physics, M&I students appear to fail to understand the basic framework of this organization of ideas, as evidenced by their near complete avoidance of the momentum principle and poor attempts at utilizing more traditional physics approaches. This suggests that the M&I curriculum must first acknowledge students' previous "traditional" physics education experiences before it can fundamentally reorganize existing and build new knowledge.

## Acknowledgments

This work was supported by National Science Foundation's Division of Undergraduate Education (DUE) 0618519.